\documentclass{aastex62}

\begin{document}

\title{Evolutionary status of selected post-AGB stars based on Gaia DR3}

\correspondingauthor{Keivan G.\ Stassun}
\email{keivan.stassun@vanderbilt.edu}

\author{M.~Parthasarathy}
\affil{Indian Institute of Astrophysics, Bangalore 560034, India}
\affil{Vanderbilt University, Department of Physics \& Astronomy, Nashville, TN  37235, USA}
\author{Marina Kounkel}
\affil{Vanderbilt University, Department of Physics \& Astronomy, Nashville, TN  37235, USA}
\author{Keivan G. Stassun}
\affil{Vanderbilt University, Department of Physics \& Astronomy, Nashville, TN  37235, USA}

\begin{abstract}
The evolutionary status of 24 post-AGB stars is presented based on Gaia DR3 data.
All 24 stars have parallaxes accurate to better than $3\sigma$ and have RUWE values $< 1.4$. Based on the Gaia DR3 distances the absolute luminosities
are derived. For 14 of the stars, the luminosities confirm their post-AGB evolutionary stage.
However, V1027 Cyg, which was previously classified as a post-AGB star, is found to have a higher luminosity; thus
it may be an evolved, massive, pulsating semi-regular variable star of type G7Ia.
For 9 of the stars, the luminosities are lower than 1000~L$_\odot$, indicating that some are 
post-HB stars and some are post-RGB stars.This
paper was completed more than four months ago but there was delay in getting it published.
\end{abstract}

\section{Introduction} \label{sec:intro}
Post-asymptotic giant branch (AGB) supergiants are stars that have recently evolved off the 
AGB but have not reached high enough temperatures to photoionize their circumstellar dust envelopes
(Parthasarathy \& Pottasch 1986, Hrivnak et al.\ 1989). 
The evolutionary stage of post-AGB supergiants is short-lived depending on the core mass (Schoenberner 1983).
 During the transition from 
the tip of the AGB to the young planetary nebula phase, these stars have
 spectral types that evolve from M to OB (Parthasarathy 1993a), and they mimic the spectra of 
supergiants because after the termination of the AGB phase of evolution they have 
a white-dwarf C–O core with a very thin extended envelope. From an analysis of IRAS data
and their multi-wavelength studies during the past 36 years have resulted in 
identifying few hundred post-AGB stars (for a list see Vickers et al.\ 2015).
 Until recently their distances were not known therefore  it was not possible
to place them on the post-AGB evolutionary tracks.
With the advent of Gaia satellite now we have accurate parallaxes for large number of post-AGB stars.
 It is now possible to derive their absolute luminosities and to compare them with the
results of the pst-AGB stellar evolutionary models (Parthasarathy et al.\ 2020, Kamath et al.\ 2022, Parthasarathy 2022,
Aoki et al.\ 2022).

In this paper we present an analysis of  Gaia DR3 data of 24 post-AGB stars.
These stars are  selected because they have accurate Gaia DR3 parallaxes, RUWE values less than 1.4 (Stassun \& Torres 2021)
and they are not known to be close binaries.
Section~\ref{sec:data} describes the selection criteria and the basic Gaia DR3 data utilized. The principal results are presented Section~\ref{sec:results} and a brief summary of the conclusions is provided in Section~\ref{sec:summary}.

\section{Selection of stars and their Gaia DR3 Data} \label{sec:data}
The Gaia DR3 data of selected post-AGB stars (selected from the list of post-AGB stars given in Vickers et al.\ 2015)
is given in Table~\ref{tab:data}.  Several stars have Gaia DR3 parallaxes comparable or less than their errors in parallaxes
and several stars have RUWE values $> 1.4$. We have not considered these stars.
The selection criteria  that we adopted  is
that they need to have accurate Gaia DR3 parallaxes ($>3\sigma$), they need to have
RUWE values less than 1.4, they are single stars (not known to be binaries) and their Gaia data has not been 
analyzed so far. With this criteria we selected 24 stars. In Table~\ref{tab:data} we also list their
$V$, $B-V$, $E(B-V)$ values and spectral types.

\begin{deluxetable}{llrrrrrrrrrl}
\tablewidth{0pt}
\tablecolumns{12}
\tablecaption{Gaia EDR3 Parallaxes and RUWE Statistics for Post-AGB Stars \label{tab:data}}
\tablehead{\colhead{Star}  &  \colhead{Name}  &  \colhead{$b$}  &  \colhead{$\pi$}  &  \colhead{$\sigma_\pi$}  &  \colhead{$d$}  &  \colhead{$\sigma_d$}  &  \colhead{$V$}  &  \colhead{$B-V$}  &  \colhead{$E(B-V)$}  &  \colhead{Sp.~Type} & \colhead{RUWE} \\
\colhead{ }  &  \colhead{ }  &  \colhead{deg.}  &  \colhead{mas}  &  \colhead{mas}  &  \colhead{pc}  &  \colhead{pc}  &  \colhead{mag.}  &  \colhead{mag.}  &  \colhead{mag.}  &  \colhead{ } & \colhead{ } }
\startdata
IRAS~01005+7910  &  ...  &  +16.6  &  0.2414  &  0.0176  &  4142.502  &  302.0217  &  10.87  &  0.17  &  0.15  &  B1.7Ibeq  &  1.037 \\
LB~3193  &  ...  &  $-$54.9  &  0.1369  &  0.0243  &  5755.1  &  691.73  &  12.70  &  $-$0.08  &  0.03  &  B5I  &  0.779 \\
IRAS~02528+4350  &  ...  &  $-$13.3  &  2.5409  &  0.0188  &  393.561  &  2.9119  &  10.68  &  0.42  &  0.44  &  A0e  &  1.234 \\
IRAS~05040+4820  &  SAO~40039  &  +04.8  &  0.3221  &  0.0145  &  3104.626  &  139.7612  &  9.74  &  0.41  &  0.50  &  A4Ia  &  1.020 \\
IRAS~05338$-$3051  &  RV~Col  &  $-$28.8  &  0.5225  &  0.0158  &  1913.876  &  57.8741  &  9.30  &  1.11  &  0.09  &  G5I  &  1.094 \\
IRAS~05381+1012  &  HD~246299  &  $-$10.6  &  1.0380  &  0.0161  &  963.39  &  14.9428  &  10.54  &  0.82  &  0.06  &  G2I  &  1.042 \\
IRAS~10456$-$5712  &  HD~93662  &  +01.5  &  0.8997  &  0.0255  &  1111.482  &  31.5025  &  6.248  &  1.706  &  0.02  &  M0III  &  0.984 \\
IRAS~11353$-$6037  &  HD~306753  &  +0.7  &  0.1692  &  0.0133  &  5910.165  &  464.5697  &  12.4  &  0.55  &  0.65  &  B5Ie  &  1.101 \\
IRAS~11385$-$5517  &  HD~101584  &  +5.94  &  0.5452  &  0.0199  &  1834.189  &  66.9486  &  7.01  &  0.39  &  ...  &  F0Iae  &  1.05 \\
IRAS~11387$-$6113  &  ...  &  +0.2  &  0.1854  &  0.0180  &  5393.743  &  523.6644  &  11.980  &  0.54  &  0.03  &  A3Ie  &  1.179 \\
IRAS~11531$-$6111  &  ...  &  +0.7  &  0.1582  &  0.0180  &  4710.18  &  423.35  &  14.99  &  1.26  &  1.29  &  B8Iae  &  0.974 \\
HD~105262  &  BD+13~2491  &  +72.47  &  0.5972  &  0.0267  &  1674.481  &  74.8638  &  7.08  &  0.0  &  0.0  &  A0Ia  &  0.8 \\
IRAS~13110$-$5425  &  HD~114855  &  +8.0  &  0.6395  &  0.0314  &  1563.722  &  76.7801  &  9.23  &  0.43  &  0.11  &  F5Ia/ab  &  1.060 \\
IRAS~14072$-$5446  &  CD$-$54~5573  &  +6.1  &  0.2072  &  0.0143  &  4826.255  &  333.0861  &  10.52  &  0.66  &  0.69  &  B7Iab  &  0.849 \\
IRAS~14488$-$5405  &  CD$-$53~5736  &  +4.5  &  0.2684  &  0.0173  &  3725.782  &  240.1492  &  10.94  &  0.72  &  0.73  &  A0Ie  &  1.327 \\
IRAS~16206$-$5956  &  CD$-$59~6142  &  $-$7.5  &  0.1815  &  0.0191  &  5020.63  &  537.73  &  9.97  &  0.34  &  ...  &  A3Iabe  &  1.07 \\
IRAS~17311$-$4924  &  Hen~1428  &  $-$9.04  &  0.2378  &  0.0201  &  3820.17  &  325.42  &  10.74  &  0.40  &  0.60  &  B3Ie  &  1.084 \\
IRAS~19410+3733  &  HD~186438  &  +7.0  &  1.0421  &  0.0196  &  959.601  &  18.0483  &  7.91  &  0.41  &  0.13  &  F3Ib  &  1.082 \\
IRAS~20004+2955  &  V1027~Cyg  &  $-$0.4  &  0.2390  &  0.0178  &  4184.100  &  311.6192  &  8.65  &  2.2  &  1.06  &  G7Ia  &  0.907 \\
IRAS~20160+2734  &  AU~Vel  &  $-$4.5  &  0.3908  &  0.0160  &  2558.854  &  104.7637  &  10.63  &  1.66  &  1.43  &  F3Ie  &  1.041 \\
IRAS~20462+3416  &  LSII+34~26  &  $-$5.75  &  0.1720  &  0.0184  &  4859.78  &  456.16  &  10.9  &  0.21  &  0.38  &  B1.5Ia/abe  &  1.235 \\
IRAS~20490+5934  &  ...  &  +9.9  &  2.0625  &  0.0155  &  484.848  &  3.6437  &  10.32  &  0.41  &  0.36  &  A3e  &  0.866 \\
IRAS~20572+4919  &  V2324~Cyg  &  +2.4  &  1.5768  &  0.0124  &  634.196  &  4.9873  &  11.63  &  1.09  &  0.83  &  F3Ie  &  0.998 \\
IRAS~21289+5815  &  ...  &  +5.2  &  1.0109  &  0.0196  &  989.218  &  19.1796  &  13.93  &  0.75  &  0.59  &  A2Ie  &  1.362 \\
\enddata 
\end{deluxetable}

\section{Results}\label{sec:results}

The Gaia DR3 parallaxes (distances) of the selected sample of stars were used to derive their
absolute luminosities. Their $V$, $B-V$ and spectral types (Table~\ref{tab:data}) were taken from SIMBAD. In order
to derive their absolute luminosities we need to take into account the interstellar and circumstellar
reddening. Many of our stars are IRAS sources and have circumstellar dust shells.

\subsection{Interstellar and circumstellar reddening}
Several stars are located at high galactic latitudes (Table~\ref{tab:data}) therefore their interstellar reddening values
are relatively low. We took a simple approach to estimate interstellar plus circumstellar reddening.
All our stars have observed $B-V$ values (Table~\ref{tab:data}). The observed $B-V$ is affected by both interstellar plus
circumstellar reddening. All of our stars have MK spectral types (Table~\ref{tab:data}). We used the calibration
between MK spectral types and corresponding intrinsic $(B-V)_0$ values  given in Allen's Astrophysical
Quantities 4th edition (Cox 2000) and Flower (1996). The difference between the observed $(B-V)$ and $(B-V)_0$ obtained from the
spectral types yields $E(B-V)$ (Table~\ref{tab:results}).

Using the following equation we derived the absolute visual magnitudes of the stars (Table~\ref{tab:results}): 
$$    M_V  = V + 5 - 5\log(d) - 3.3 E(B-V) . $$

\begin{figure}[!ht]
    \centering
    \includegraphics[width=0.95\linewidth,trim=70 65 70 70,clip]{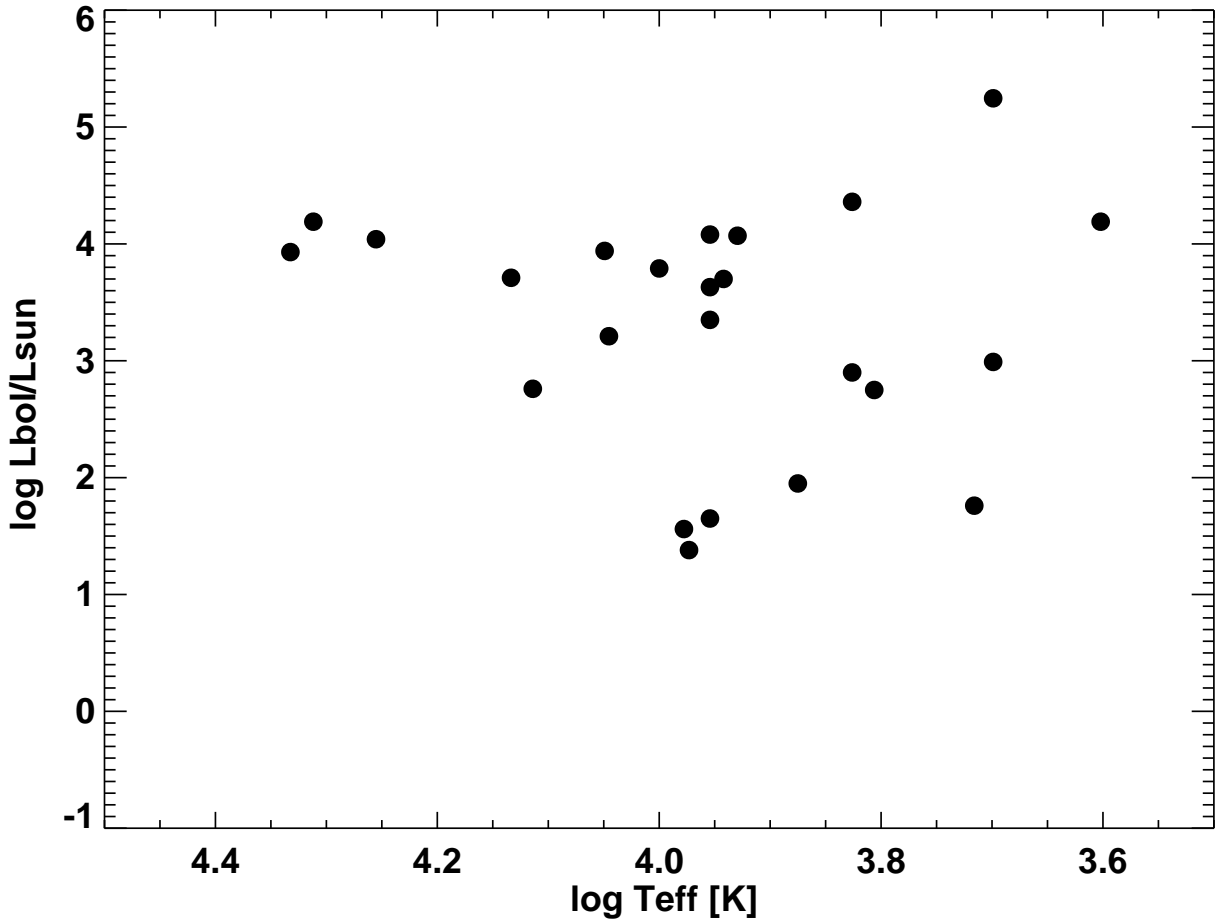}
    \caption{HR Diagram positions of the sample stars.}
    \label{fig:hrd}
\end{figure}

\begin{deluxetable}{lrrrr}
\tablewidth{0pt}
\tablecolumns{5}
\tablecaption{Absolute luminosities of the stars\label{tab:results}}
\tablehead{\colhead{Star} &  \colhead{$M_V$} & \colhead{$M_{\rm bol}$} &  \colhead{$T_{\rm eff}$} &  \colhead{$\log L/L_\odot$} \\
\colhead{ } &  \colhead{mag.} & \colhead{mag.} &  \colhead{K} &  \colhead{dex} }
\startdata
IRAS~01005+7910  &  $-$3.503  &  $-$5.083  &  21500  &  3.93 \\
LB3193  &  $-$1.2  &  $-$2.15  &  13000  &  2.76 \\
IRAS~02528+4350  &  +1.253  &  +0.843  &  9500  &  1.56 \\
IRAS~05040+4820  &  $-$4.37  &  $-$4.50  &  8750  &  3.70 \\
RV~Col  &  $-$2.407  &  $-$2.737  &  5000  &  2.99 \\
IRAS~05381+1012  &  +0.423  &  +0.273  &  5200  &  1.76 \\
IRAS~10456$-$5712  &  $-$4.296  &  $-$5.746  &  4000  &  4.19 \\
IRAS~11353$-$6037  &  $-$3.573  &  $-$4.523  &  13600  &  3.71 \\
IRAS~11385$-$5517  &  $-$5.33  &  $-$5.44  &  8500  &  4.07 \\
IRAS~11387$-$6113  &  $-$3.362  &  $-$3.642  &  9000  &  3.35 \\
IRAS~11531$-$6111  &  $-$2.632  &  $-$3.292  &  11100  &  3.21 \\
HD~105262  &  $-$3.928  &  $-$4.328  &  9000  &  3.63 \\
IRAS~13110$-$5425  &  $-$2.104  &  $-$2.134  &  6400  &  2.75 \\
IRAS~14072$-$5446  &  $-$5.175  &  $-$5.455  &  9000  &  4.08 \\
IRAS~14488$-$5405  &  $-$4.325  &  $-$4.735  &  10000  &  3.79 \\
IRAS~16206$-$5956  &  $-$4.689  &  $-$5.099  &  11200  &  3.94 \\
IRAS~17311$-$4924  &  $-$4.15  &  $-$5.73  &  20500  &  4.19 \\
IRAS~19410+3733  &  $-$2.495  &  $-$2.515  &  6700  &  2.90 \\
IRAS~20004+2955  &  $-$7.956  &  $-$8.376  &  5000  &  5.246 \\
IRAS~20160+2734  &  $-$6.129  &  $-$6.16  &  6700  &  4.36 \\
IRAS~20462+3416  &  $-$3.787  &  $-$5.367  &  18000  &  4.04 \\
IRAS~20490+5934  &  +0.704  &  +0.604  &  9000  &  1.65 \\
IRAS~20572+4919  &  $-$0.12  &  $-$0.13  &  7500  &  1.95 \\
IRAS~21289+5815  &  +1.578  &  +1.298  &  9400  &  1.38 \\
\enddata
\end{deluxetable}

\subsection{Luminosities}
To calculate the absolute bolometric magnitudes $M_{\rm bol}$ we used the bolometric corrections (BC)
given in Allen's Astrophysical Quantities (4th edition) (Cox 2000). The derived $M_{\rm bol}$ values and $\log (L/L_\odot)$
values are given in Table~\ref{tab:results}. For several stars  in this sample $T_{\rm eff}$ values are available
in the literature (based on the analysis of their spectra). The references for $T_{\rm eff}$ values are  given in the
Section~\ref{sec:notes} notes. We have also used spectral types, $(B-V)_0$ and $T_{\rm eff}$ calibration given in
Allen's Astrophysical Quantities (4th edition) (Cox 2000). $T_{\rm eff}$ values are not used in deriving the
absolute luminosities. $T_{\rm eff}$ values are used only to examine the location of the stars in the H-R diagram (Figure~\ref{fig:hrd}).

\subsection{Uncertainties}
The errors in parallaxes and distances are given Table~\ref{tab:data}. The errors in $E(B-V)$ are of the order of  $\pm$0.05
to $\pm$0.10.  The errors in derived $M_V$ values are on the average less than 0.1. The errors in derived
absolute luminosities $\log (L/L_\odot)$ of the order 0.10 dex. The errors in $T_{\rm eff}$ values are of the order of 500~K.

\subsection{Notes on individual sources}\label{sec:notes}

\subsubsection{IRAS 01005+7910}
It is a high galactic latitude  carbon-rich hot post-AGB star. Zhang \& Kwok (2011) found fullerene C60 in the 
Spitzer/IRS mid-IR spectrum of this star. Iglesias-Groth et al.\ (2013) found absorption bands at 9577\AA\ and 9632\AA\ which are due
to  fullerene C60+ cation. Klochkova et al.\ (2002) from an analysis of high resolution optical spectrum of this star
derived $T_{\rm eff}$ = 21,500~K and found it to be carbon-rich. Thus the photosphere and circumstellar matter of this star is carbon-rich.
It appears that the star has evolved from a carbon-rich AGB star to carbon-rich post-AGB stage and it may evolve
into a [WC] type planetary nebula (PN) (Parthasarathy 1999). Its luminosity  and $T_{\rm eff}$ (Table~\ref{tab:results}) confirms that it is a
hot post-AGB star. It may soon evolve into a young PN similar to Hen 1357 (SAO 244567) (Parthasarathy et al.\ 1993).

\subsubsection{LB 3193}
It is a high galactic  latitude hot metal-poor star. McCausland et al.\ (1992) analyzed high resolution optical
spectrum of this star and they classified it as a galactic halo hot post-AGB star. They found underabundance of metals
and severe underabundance of carbon. The star seems to have left the AGB stage of evolution before the third dredge up.
They derived $T_{\rm eff}$ = 13,000~K, $\log g = 2.2$ and [Fe/H] = $-1.0$. Quin \& Lamers (1992) analyzed UV (IUE) spectrum
of this star and derived $T_{\rm eff}$ = 14,000~K and $E(B-V) = 0.03$.  It is not an IRAS source hence there is no circumstellar
dust envelope. From Gaia DR3 data we derive the absolute luminosity $\log (L/L_\odot) = 2.76$. Its luminosity
is very low when compared with the luminosity of a typical post-AGB star. It may be a post-HB star or a AGB-manque star.

\subsubsection{IRAS 02528+4350}
It is classified as a high galactic latitude post-AGB star of spectral type A0e (Suarez et al.\ 2006).
Fujii et al.\ (2002) made $BVRIJHK$ photometric observations. From Gaia DR3 data (Tables \ref{tab:data} and \ref{tab:results}) we find that
its absolute luminosity is only  36.3 L$_\odot$. It is not a post-AGB star. It may be a post-HB star or a
high galactic latitude A-type  young star with  circumstellar dust shell/disk.
 To further understand its evolutionary status an analysis
of its high resolution optical spectrum is needed. It may be a normal A-type dwarf.

\subsubsection{IRAS 05040+4820 (SAO 40039)}
On the basis of IRAS colors and far-IR flux distribution, high galactic latitude and A4 Ia spectral type
Parthasarathy (1993b) classified it as a post-AGB star. Fujii et al.\ (2002) made $BVRIJHK$ photometry
and modeled the SED including the near and far-IR (IRAS) data and concluded that it is a low-mass
post-AGB star. Parthasarathy et al.\ (2005) made $BVRI$ polarization measurements and concluded that the
circumstellar dust shell is asymmetric. Rao et al.\ (2011) find that the this star may be mildly He-rich.
However the $T_{\rm eff}$ used by them in the analysis of spectrum may be relatively low. Fujii et al.\ (2000)
from SED find $T_{\rm eff}$ = 8750~K. Klochkova et al.\ (2015) presented a spectral atlas. Parthasarathy et al.\ (2020)
obtained K-band spectrum and find no He~I and H2 lines. They find the Brackett Gamma to be a broad absorption line.
The absolute luminosity derived from the Gaia DR3 data (Tables \ref{tab:data} and \ref{tab:results}) is in good agreement with the
post-AGB evolutionary track of a low-mass star.

\subsubsection{RV Col (IRAS F05338-3051) (CD-30 2512)}

It is a high galactic latitude G5 semi-regular (SRD) variable with a period of 105.7 days.
The variations in light are up to one magnitude.  Arkhipova et al.\ (2011) studied the photometric variability
and considered it as a post-AGB candidate. The luminosity derived from Gaia DR3 data (Tables \ref{tab:data} and \ref{tab:results})
shows that it is not a post-AGB star. It may be a post-RGB star. Detailed chemical composition study
of this star from an analysis of high resolution spectra is needed. Large amplitude light variations
also indicate that it is not a post-AGB star. However the presence circumstellar dust shell
indicates that it suffered mass-loss in the recent past.

\subsubsection{IRAS 05381+1012}
It is a high galactic latitude  G2I star.  Pereira \& Roig (2006) analyzed high resolution spectrum of this star.
They classified it as a post-AGB star and estimated a distance of 2700~pc. They  
have derived $T_{\rm eff}$ = 5200~K, $\log g  = 1.0$ and [Fe/h] = $-0.8$,  and rotational velocity of 40~km/sec. We find Gaia DR3 data
(Tables \ref{tab:data} and \ref{tab:results}) that its distance is 963.39~pc and $\log (L/L_\odot) = 2.00$. Its luminosity
is too low hence it is not a post-AGB star. It may be post-RGB star. The presence of circumstellar dust shell,
low metallicity, high galactic latitude, raid rotation makes this an interesting object for further study.
Kounkel et al.\ (2018, 2019) derived $T_{\rm eff}$ = 5493~K, $\log g = 3.26$ and  rotational velocity of 50.89~km/sec and radial velocity
of 43.280~km/sec.

\subsubsection{IRAS 10456-5712}
It is a relatively bright star. In SIMBAD its spectral type is K5/M0III. There is no spectroscopic
study of this star.  On the basis of IRAS colors and far-IR flux distribution     it is classified as a post-AGB star.
It may have cold circumstellar dust disk. From Gaia DR3 data (Tables \ref{tab:data} and \ref{tab:results}) 
we derive a distance of 1111.482~pc and $\log (L/L_\odot) = 4.194$. It may be a post-AGB star. High resolution
spectroscopic study is needed to further understand its evolutionary status.

\subsubsection{IRAS 11353-6037 (HD 306753)}

It is a hot  (B5Ie) post-AGB star (Suarez et al.\ 2006). The luminosity (Table \ref{tab:results}) suggests that it is a post-AGB star.
There is no detailed study of this star.

\subsubsection{IRAS 11385-5517 (HD 101584)}
On the basis of IRAS colors, far-IR flux distribution and circumstellar dust shell parameters 
Parthasarathy \& Pottasch (1986) classified it as a post-AGB F0Ia supergiant similar to HD 161796.
The near and far-IR flux distribution shows that it is having hot and cold dust circumstellar  dust shell/disk.
Olofsson et al.\ (2021) classify it as post-RGB star. Sivarani et al.\ (1999) from an analysis high resolution spectra
derived $T_{\rm eff}$ = 8500~K, $\log g = 1.5$ and [Fe/H] = 0.0. They found several emission lines and P-Cygni profiles indicating the presence of
stellar wind, very extended envelope and circumstellar disk. Diaz et al.\ (2007) found radial velocity variations and they
derived a period of 144 days. It shows bipolar flows and complex circumstellar envelope/disk (Olofsson et al.\ 2021). 
Olofsson et al.\ (2021) find heavy element Rydberg transition line emission using high angular resolution ALMA
observations.  From the analysis Gaia DR3 data (Tables \ref{tab:data} and \ref{tab:results}) we confirm that it is a post-AGB star
with $\log (L/L_\odot) = 4.072$. The Gaia parallax yields a distance of 1834.189~pc. Its luminosity (Table \ref{tab:results}) indicates
that the progenitor main-sequence star may be of  2 to 2.5 M$_\odot$.

\subsubsection{HD 105262}
It was discovered by Abt (1996) to be a metal-poor post-AGB star with large proper motion. Abt (1996) concluded
that its spectrum is similar to that of post-AGB star HR 4049. It is a very high galactic latitude B9I star.
Reddy et al.\ (1996) and Arentsen et al.\ (2019) analyzed the spectrum of this star and find that
$T_{\rm eff}$ = 8225~K, $\log g = 1.4$, [Fe/H] = $-1.94$. Its Gaia DR3 data reveals it is at a distance of 1674.481~pc. Its luminosity
(Table \ref{tab:results}) confirms that it is a post-AGB star.  It is not an IRAS source. There seems to be no circumstellar dust shell.

\subsubsection{IRAS 13110-5425 (HD 114855) (V956 Cen)}
It is a high galactic latitude F5Ia/ab star. The luminosity derived from Gaia DR3 data (Tables \ref{tab:data} and \ref{tab:results})
is significantly lower than that of a post-AGB star. It may be a post-RGB star.  There is no high resolution
spectroscopic study of this star.

\subsubsection{IRAS 16206-5956 (SAO 243756) (CD-59 6142)}
Parthasarathy et al.\ (2000) presented low resolution spectrum of this star and classified it as a high galactic latitude
A3Iabe post-AGB star. Gauba \& Parthasarathy (2003, 2004) find variability in the UV (IUE) spectrum. They derived $T_{\rm eff}$ 
= 11,200~K, $\log g = 2.3$. The luminosity derived from Gaia DR3 data (Tables \ref{tab:data} and \ref{tab:results}) confirms that it is a post-AGB star.

\subsubsection{IRAS 17311-4924 (Hen 1428)}
On the basis of IRAS colors, far-IR flux distribution and circumstellar dust shell parameters Parthasarathy \& Pottasch (1989)
were the first to classify it as a high galactic latitude hot post-AGB star. Gauba \& Parthasarathy (2003, 2004),
Sarkar et al.\ (2005) and Mello et al.\ (2012) studied this star and derived $T_{\rm eff}$ = 20,500~K, $\log g = 2.35$. The luminosity derived
from Gaia DR3 data (Tables \ref{tab:data} and \ref{tab:results}) confirms that it is a post-AGB star.

\subsubsection{IRAS 19410+3733 (HD 186438) (LSII +37 1)}
The luminosity derived from Gaia DR3 data (Tables \ref{tab:data} and \ref{tab:results}) indicates that it is not a post-AGB star.
It may be a post-RGB star.

\subsubsection{IRAS 20004+2955 (V1027 Cyg) (HD 333385)}
It is in the galactic plane. It is a semiregular variable star of spectral type G7Ia.
Arkhipova et al.\ (2016) studied the photometric and spectral variability. The luminosity derived from
Gaia DR3 data  (Tables \ref{tab:data} and \ref{tab:results}) indicates that it is not a post-AGB star. It is most likely a massive star.

\subsubsection{IRAS 20462+3416  (LS II +34 26)}
On the basis of IRAS colors, far-IR flux distribution, and cicumstellar dust shell parameters Parthasarathy (1993a)
discovered it to be a hot (B1.5Ia) post-AGB star.  Garcia-Lario et al.\ (1977) analyzed high resolution optical and UV (IUE)
spectrum of this star and derived $T_{\rm eff}$ = 18000~K. The presence of low excitation nebular emission lines was  first
reported by Parthasarathy (1994) and he has concluded that it is a rapidly evolving hot post-AGB star
similar to SAO 244567 (Parthasarathy et al.\ 1993).  The luminosity derived from Gaia DR3 data
(Tables \ref{tab:data} and \ref{tab:results}) confirms that it is a post-AGB star. Parthasarathy et al.\ (2020) from the K-band spectrum of this star
find the presence of H2 lines at 2.12, 2.22, and 2.247~$\mu$m. They have also find the Br~$\gamma$ emission line in the spectrum.
Gledhill \& Forde (2015) made a detailed study of the  2~$\mu$m to 2.4~$\mu$m spectrum of this star and found that
Br~$\gamma$ emission is spatially extended.

\subsubsection{IRAS 20490+5934}
The luminosity (Tables \ref{tab:data} and \ref{tab:results}) indicates that it is not a post-AGB star.

\subsubsection{IRAS 20572+4919  (V2324 Cyg)}
The luminosity (Tables \ref{tab:data} and \ref{tab:results}) indicates it is not a post-AGB star.

\subsubsection{IRAS 21289+5815}
The luminosity (Tables \ref{tab:data} and \ref{tab:results}) indicates it is not a post-AGB star.

\section{Conclusions}\label{sec:summary}

Based on the Gaia DR3 parallaxes of 24 selected post-AGB stars we derived their absolute luminosities.
We find 14 of them have luminosities that confirm their post-AGB status.  HD 101584
which was originally classified as a  post-AGB star (Parthasarathy \& Pottasch 1986); it
 is indeed a post-AGB star and not a
a post-RGB star. LS II +34 26, SAO 40039, SAO 243756 and Hen 1428 are also confirmed
to be hot  post-AGB stars from this study. 

Nine of the stars have luminosities
$\log (L/L_\odot$) much less than 3.5. LB 3193 which was classified as a  high galactic latitude,
 galactic halo metal-poor hot post-AGB star is found to be only 575~L$_\odot$. It may be a
 post-HB star or a AGB-manque star.
Some of the less luminous stars may be post-HB stars and some may be post-RGB stars.
Some may be misclassified as post-AGB stars. 

Finally, V1027 Cyg is found to have much higher
luminosity than that of a typical post-AGB star. V1027 Cyg seems to be a massive, very luminous
semiregular variable G7Ia supergiant.

\section*{References}

Abt,H., 1996, PASP 108, 849

Aoki,W., Matsuni, T., Parthasarathy,M., 2022, PASJ (in press)

Arentsen,A., et al., 2019, A\&A 627, 138

Arkhipova, V.P., et al., 2011, AstL 37, 635

Arkhipova, V.P., et al., 2016, AstL 42, 756

Cox,A.N., 2000, Allen's Astrophysical Quantities (4th edition)

Diaz,F., et al., 2007, IAU Symp. 240, 127

Flower,P.J., 1996, ApJ 469, 355

Fujii,T., Nakada,Y., Parthasarathy,M., 2002, A\&A 385, 884

Garcia-Lario,P., Parthasarathy,M., De -Martino,D., et al., 1997, A\&A 326, 110

Gauba,G., \& Parthasarathy,M., 2003,  A\&A 407, 1007

Gauba,G., Parthasarathy,M., 2004, A\&A 417, 201

Hrivnak, B.J., et al. 1989, ApJ 346, 265

Gledhill, T,M., \& Forde, K.P., 2015, MNRAS 447, 1080

Iglesias-Groth,S., et al., 2013,  ApJ 776, L2

Kamath,D., et al., 2022, ApJ, 927, L13

Klochkova,V.G., 2002, A\&A 392, 143

Klochkova,V.G., et al., 2015, AstBu., 70, 99

Kounkel,M., et al., 2018, AJ, 156, 84

Kounkel,M., et al., 2019, AJ 157, 196

McCausland, R.J.H., et al., 1992, ApJ 394, 298

Mello, D.R.C., et al., 2012, A\&A 

Miller-Bertolami,M.M., 2016, A\&A 588, 25

Olofsson,H., et al., 2021, A\&A 651, 35

Parthasarathy,M., 1993a, ApJ 414, L109

Parthasarathy,M., 1993b, ASP Conf. Series 45, 173

Parthasarathy,M., 1994, ASP Conf.Series 60, 261

Parthasarathy,M., 1999, IAU Symp. 191, 475

Parthasarathy,M., 2022, RNAAS 6, 33

Parthasarathy,M., et al., 1993, A\&A 267, L19

Parthasarathy,M., Pottasch, S.R., 1986, A\&A 154, L16

Parthasarathy,M., Pottasch,S.R., 1989, A\&A 225, 521

Parthasarathy, M., Jain, S.K., Sarkar, G., 2005, AJ 129, 2451

Parthasarathy,M., Matsuno,T., Aoki,W., 2020, PASJ 72, 99

Parthasarathy,M., Muthumariappan,C., Muneer,S., 2020, Ap\&SS 365, 127

Parthasarathy,Vijapurkar,J., Drilling, J.S., 2000, A\&AS 145, 269

Pereira,C.B., \& Roig, F., 2006,  A\&A 452, 571

Pereira,C.B., Miranada,L.F., 2007, A\&A 462, 231

Quin, D.A., \& Lamers, H.J.G.L.M., 1992, A\&A 260, 261

Rao,S.S., et al., 2011, ApJ 737, L7

Reddy, B.E., Parthasarathy,M., Sivarani, T., 1996, A\&A 313, 191

Sarkar,G., Parthasarathy,M., Reddy,B.E., 2005, A\&A 431, 1007

Schoenberner,D., 1983, ApJ 272, 708

Sivarani,T., Parthasarathy,M., Garcia-Lario, P., et al., 1999, A\&AS 137, 505

Stassun,K.G., \& Torres, G., 2021, ApJ 907, L33

Suarez,O., et al., 2006, A\&A 458, 173

Vickers, S.B., et al., 2015, MNRAS 447, 1673

Zhang,Y.,  Kwok,S., 2011, ApJ 730, 126

\end{document}